# Parameterization of photoinhibition for phytoplankton

**Mohammad M. Amirian [1] ✉, Zoe V. Finkel [2], Emmanuel Devred [3] & Andrew J. Irwin [1]**

Mathematical models of photosynthesis–irradiance relationships in phytoplankton are used to compute integrated water-column photosynthetic rates and predict primary production. Models typically ignore an important phenomenon observed in most experiments: photosynthetic rate remains constant over a range of irradiances before declining due to photoinhibition. Here we develop an approach that captures both photoinhibition and this plateau. We test six new models of photoinhibition and ten more photoinhibition models from the literature against a database of 1808 photosynthesis–irradiance curves exhibiting photoinhibition. The best model improves adjusted $R^2$ by 6–37% and reduces RMSE by 15–70% compared to existing models. In the best model, photoinhibition is phenomenologically described by multiplication by a saturating function of the reciprocal of irradiance, simplifies to the widely-used Jassby & Platt photosynthesis–irradiance curve in the absence of photoinhibition, and only requires one new parameter. This photoinhibition parameter identifies the onset of photoinhibition and is the rate of decrease in photosynthetic rate at that irradiance. Simulations show that while parameter values vary up to 40% across existing models, particularly with and without photoinhibition, our representation of photoinhibition does not affect the interpretation or numerical values of parameters compared to the corresponding model without photoinhibition.

Predictions of photosynthetic rates of phytoplankton are used in estimates of water-column integrated primary production over large areas[1–4] and ecosystem models[5,6]. The photosynthesis–irradiance (PI) curve is an important mathematical model underlying these computations. Parameter values in the PI curve are generally determined from statistical analysis of measured rates of photosynthesis from light incubation experiments. Many formulations of PI curves have been developed and several competing models are commonly used[7,8]. A comprehensive list of existing PI curve models is provided in Table S1 among which three commonly used formulations for photosynthetic rate, P (mol C (mg chl a)$^{-1}$ h$^{-1}$), as a function of irradiance, I (μmol photons m$^{-2}$ s$^{-1}$), are

$$P = P_{max} \tanh(\alpha I/P_{max}), \quad (1a)$$

$$P = P_{max}(1 - \exp(-\alpha I/P_{max})), \text{ and} \quad (1b)$$

$$P = P_{max} I/(I + P_{max}/\alpha), \quad (1c)$$

where the photosynthetic efficiency at low light is described by α and the photosynthetic capacity (maximum rate of photosynthesis) is $P_{max}$[9–11]. Photosynthetic rate is typically measured by the incorporation of radiolabeled bicarbonate into organic biomass and reported normalized to chlorophyll *a* content as a proxy for phytoplankton biomass. The derived parameter $P_{max}/\alpha$, commonly referred to as $I_k$ or $E_k$, is frequently used to approximate the irradiance at which photosynthesis becomes light-saturated. In this study, we refer to this parameter as $I_\alpha$ to emphasize its dependence on the photosynthetic efficiency parameter at low light level, α. The parameter α can be written as the product of the chlorophyll-normalized absorption cross-section, $a^*$ (m$^2$ (mg chl a)$^{-1}$), the quantum yield of photosynthesis, ϕ, equal to the ratio of the mol of organic C or $O_2$ produced to the mol of photons absorbed, and a unit conversion constant. Photosynthetic rate generally increases nearly linearly as irradiance increases under low light, then reaches a maximum rate. At relatively high irradiance, increasing light reduces the photosynthetic rate. The formulas in Eq. (1) do not describe a decrease in photosynthetic rate with increasing irradiance, so if a decrease is observed in data, there will be bias in the estimated parameters.

Photoinhibition is a general term for the reduction of photosynthetic rate with increasing irradiance at relatively high irradiance, but it is a composite phenomenon arising from many distinct processes. Photosystem

[1]Department of Mathematics & Statistics, Dalhousie University, Halifax, NS, Canada. [2]Department of Oceanography, Dalhousie University, Halifax, NS, Canada. [3]Fisheries and Oceans Canada, Bedford Institute of Oceanography, Dartmouth, NS, Canada. ✉e-mail: m.amirianmatlob@dal.ca

II is susceptible to light-dependent photoinactivation which can be countered by repair with associated metabolic and opportunity costs[12,13]. The magnitude of the observed decrease in photosynthetic rate can increase with the duration of exposure to high light as this repair capacity is overwhelmed. Phytoplankton can acclimate to incident irradiance by remodeling their photosynthetic apparatus, in particular the amount of chlorophyll *a*, accessory and photoprotective pigments, altering the photosynthetic efficiency and quantum yield with non-photochemical quenching, adjusting maximum photosynthetic capacity and the irradiance at the onset of photoinhibition[14]. Individual taxa in the community will vary in their capacity to use each of these mechanisms. Mechanisms such as photoinactivation operate at all irradiances, including very low levels, while other processes and the aggregate effect of photoinhibition are typically only observed at irradiances exceeding the saturating irradiance. The observed photosynthesis–irradiance response of a natural community of phytoplankton is thus a complex synthesis of many mechanisms, the taxonomic composition of the community, and the history of light and other conditions and resources needed for photosynthesis and acclimation (e.g., temperature, nutrient concentrations).

To simplify this complexity, a variety of simple empirical models to parameterize photoinhibition have been developed (Table 1). The most widely used approach incorporates an exponential function to represent the photoinhibition component. This exponential term is typically multiplied with light-saturating models (i.e., those listed in Table S1 and Eq. 1) to capture the shape of the PI curve in the presence of photoinhibition. The use of an exponential function to model photoinhibition was first introduced by Steele in 1962[15] and has since been adapted into several empirical formulations to better fit observed data and enable estimation of the photoinhibition parameter, β (Table 1). A commonly used formulation for modeling the PI curve with photoinhibition is the product of Eq. (1b) and an exponential decay term, $\exp(-\beta I/P_s)$:

$$P = P_s(1 - \exp(-\alpha I/P_s)) \exp(-\beta I/P_s) \tag{2}$$

where β is a photoinhibition parameter with the same units as α, and $P_{max}$ is replaced with $P_s$ to indicate that it no longer represents the empirically observed maximum photosynthetic rate, but rather the *theoretical* maximum photosynthetic rate[16,17]. To estimate the true photosynthetic capacity ($P_{max}$), one must first estimate $P_s$ from data and then numerically derive $P_{max}$ using a formula such as the ones provided in Table S2. The estimate of $P_{max}$ is susceptible to error propagation and bias from the estimates of parameters α and β potentially leading to bias in estimates of primary production in the water column.

In commonly used PI curve formulations that incorporate photoinhibition (e.g., Eq. 2) photodamage begins at the lowest irradiances, gradually overwhelms the advantages of increasing irradiance, and eventually photosynthetic rate decreases exponentially with increasing irradiance. The PI curve has a maximum achieved photosynthetic rate at a single irradiance and photosynthetic rate then immediately decreases with further increases in irradiance (Fig. 1, top panel). Phytoplankton are typically able to achieve their maximum photosynthetic rate over a wide range of irradiances, which we call a plateau in the PI curve (Fig. 1, bottom panel)[18,19], but Eq. (2) is not able to describe this plateau. All models without photoinhibition incorporate a plateau, but only two photoinhibition models in the literature describe a plateau. One approach is a piecewise function with a constant plateau[20], but there is rarely enough data to accurately determine the irradiance where the plateau joins the decreasing part of the curve. A second approach incorporates an additional shape parameter, even when photoinhibition is not present[21].

We develop a PI curve model that accurately captures the plateau region of the PI curve in the presence of photoinhibition, while seamlessly simplifying to an established PI formulation when photoinhibition is absent. The model is designed so that photoinhibition does not alter the interpretation of other key parameters, all parameters can be reliably estimated from typical datasets with minimal propagation of error, and to have a simple geometric interpretation of the photoinhibition parameter. We propose to model the plateau in photosynthetic rate by multiplying the saturating function describing light-limited and light-saturated photosynthesis (Eq. 1) by a saturating function evaluated at the reciprocal of irradiance. The effect of using the reciprocal irradiance is to reduce photosynthetic rate as irradiance increases and the effect of the saturating function is to enable a plateau and the simplification to the original model in the absence of photoinhibition. Our new phenomenological model is

$$P = P_{max} \tanh\left(\frac{\alpha I}{P_{max}}\right) \tanh\left(\left(\frac{P_{max}}{\beta I}\right)^{\gamma}\right), \tag{3}$$

**Table 1 | Established and new photoinhibition models formulated with the parameters $P_{max}$, $I_\alpha$, $I_\beta$, $P_s$, $I^s_\alpha$, $I^s_\beta$ and a shape parameter γ**

| Name | Equation |
|---|---|
| Established 3 parameter model | |
| Steele[15] | $P_s\left(\frac{I}{I^s_\alpha}\right)e^{1-\left(\frac{I}{I^s_\alpha}\right)}$ |
| Established 4 parameter models | |
| Peeters and Eilers[28] | $P_s\left(I/I^s_\alpha\right)\left(1+\frac{I}{I^s_\alpha}+\frac{I^2}{I^s_\alpha I^s_\beta}\right)^{-1}$ |
| Platt et al.[16] | $P_s\left(1-e^{-I/I^s_\alpha}\right)e^{-I/I^s_\beta}$ |
| Neale and Richerson[20] | $P_s \tanh\left(I/I^s_\alpha\right)e^{-I/I^s_\beta}$ |
| Baly*[9] | $P_s\left(\frac{I}{I+I^s_\alpha}\right)e^{-I/I^s_\beta}$ |
| Smith*[32] | $P_s\left(\frac{I}{\sqrt{I^2+(I^s_\alpha)^2}}\right)e^{-I/I^s_\beta}$ |
| Blackman*[33] | $P_{max}\begin{cases} I/I_\alpha, & \text{if} I<I_\alpha \\ 1, & \text{if} I_\alpha \le I<I_\beta \\ 1-\left(I-I_\beta\right)/I_\beta, & \text{if} I_\beta \le I \end{cases}$ |
| Established 5 parameter models (with shape parameter $0<\gamma<1$ *for* models marked with[a]) | |
| Bannister*[26] | $P_s\left(\frac{I}{\sqrt[\gamma]{I^\gamma+(I^s_\alpha)^\gamma}}\right)e^{-I/I^s_\beta}$ |
| Prioul and Chartier*[25] [a] | $\frac{P_s}{2\gamma}\left(\tilde{I}-\sqrt{\tilde{I}^2-4\gamma I/I^s_\alpha}\right)e^{-I/I_\beta}$ where $\tilde{I}=1+\left(I/I^s_\alpha\right)$ |
| Fasham and Platt[21][a] | $\frac{P_s}{2\gamma}\left(\tilde{I}-\sqrt{\tilde{I}^2-4\gamma I/I^s_\alpha}\right)$ where $\tilde{I}=1+\left(\frac{I}{I^s_\alpha}\right)\left(\gamma+(1-\gamma)e^{I/I^s_\beta}\right)$ |
| New models | |
| Amirian | $P_{max}\tanh\left(I/I_\alpha\right)\tanh\left(\left(I_\beta/I\right)^{\cosh^2(1)}\right)$ |
| exp-tanh | $P_{max}\left(1-e^{-I/I_\alpha}\right)\tanh\left(\left(I_\beta/I\right)^{\cosh^2(1)}\right)$ |
| tanh-exp | $P_{max}\tanh\left(I/I_\alpha\right)\left(1-e^{-I_\beta/I}\right)$ |
| exp-exp | $P_{max}\left(1-e^{-I/I_\alpha}\right)\left(1-e^{-I_\beta/I}\right)$ |
| tanh-tanh-γ | $P_{max}\tanh\left(I/I_\alpha\right)\tanh\left(\left(I_\beta/I\right)^\gamma\right)$ |
| exp-tanh-γ | $P_{max}\left(1-e^{-I/I_\alpha}\right)\tanh\left(\left(I_\beta/I\right)^\gamma\right)$ |

Established models are identified by reference and new models are identified by name (Amirian) or functional form (e.g., exp-tanh). Models marked with a star (*) did not originally incorporate photoinhibition and have been modified.
In many models $P_{max}$ is replaced with $P_s$ as this factor does not represent the photosynthetic capacity; irradiance parameters have an s superscript since they are defined using $P_s$ instead of $P_{max}$, and $P_{max}$ must be computed from $P_s$ as shown in Table S2. All models are fit with a constant intercept, R, which was omitted from the table.

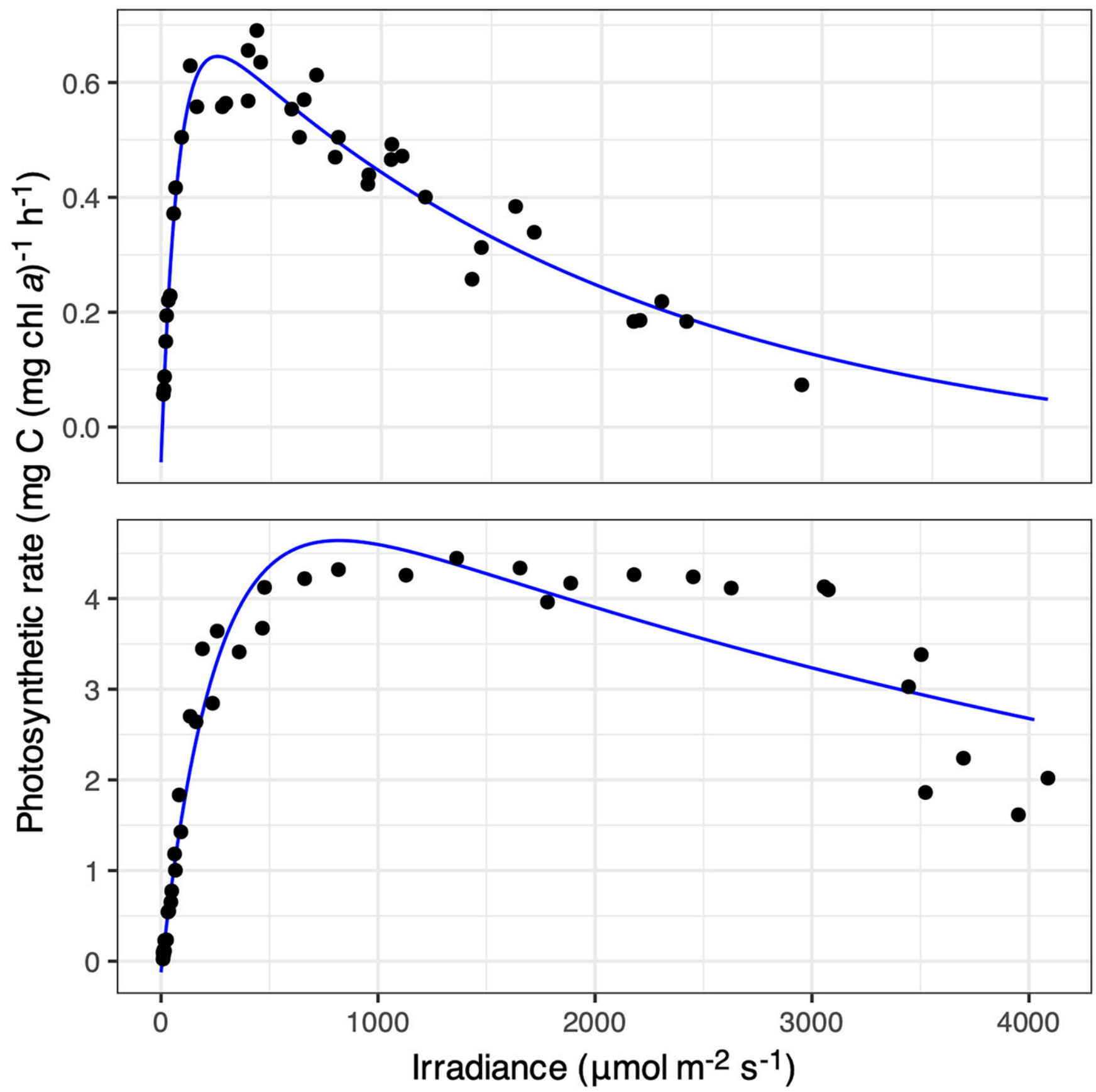

**Fig. 1 | Data from two photosynthesis–irradiance experiments exhibiting different degrees of photoinhibition.** Points are data[21], lines are model fits (Eq. 2).

where β and $I_\beta = P_{max}/\beta$ are parameters describing the rate and onset of photoinhibition analogous to the efficiency parameter, α, and saturation irradiance, $I_\alpha$, with the same units as those parameters (Fig. S8). We included a dimensionless shape parameter, γ, for additional flexibility, but our analysis will support a fixed value for γ. The second tanh factor is approximately 1 at sub-saturating irradiance so that the effect of photoinhibition is absent at low irradiance (Fig. S8, right panel). Here we test this model and several variations that combine our idea with existing models together with several models from the literature and show the utility of our new model. We evaluate 16 models statistically on a large data set of PI experiments, determine the best model and importance of the plateau, and compare the interpretation of PI parameters across the models.

## Materials and methods

### Model selection and development

We developed a set of models to test based on models in the literature and our ideas (Table 1). First, we selected the most widely used photoinhibition model that combines the exponential model of light-saturated photosynthesis and the exponential model of photoinhibition (Eq. 2). We added a variety of influential photoinhibition models from the literature selected based on their prevalence in the literature and the diversity of algebraic formulation. To increase the pool of models, we created some new models that combined existing models of light-saturated photosynthesis and exponential photoinhibition that we did not find in the literature. We included our new model (Eq. 3) and five variations incorporating our key idea of using reciprocal irradiance with two light-saturating models (Eq. 1a, 1b) and using those two saturating functions to describe the onset of photoinhibition, plus an optional shape parameter (Supplemental Methods) All models included a constant intercept, R, to allow for error in the measurement from the dark bottle and respiration. The models have many similarities although they vary in algebraic complexity and the number (3 to 5) of parameters.

### PI data sets

We compiled PI curves collected from 1973–2022 by scientists at Fisheries and Oceans (DFO) Canada. To ensure data integrity, we implemented quality control measures to remove errors found in electronic data tables and cruise reports (removing duplicates, combining data from multiple sources, correcting errors, ensuring consistent units). The dataset consists of 3641 PI incubation experiments gathered from 1304 locations, predominantly situated in the northern hemisphere (Fig. S1). Photosynthesis–irradiance curves were obtained from phytoplankton community samples gathered in Niskin bottles at two depths (typically near surface 0–10 m and sub-surface 10–50 m), returned to the deck of the ship, spiked with $^{14}$C-labeled bicarbonate, and incubated for 2–6 h (typically 4 h) under a 150 W floodlight. A dark bottle was used as a blank. The temperature inside the incubator was controlled by pumping seawater through the incubator. Methods changed little over the collection period; for more details, see Irwin et al.[22]. Photosynthetic rates and the photosynthetic capacity (mg C (mg chl-*a*)$^{-1}$ h$^{-1}$) are normalized to biomass quantified as chlorophyll *a* concentration (mg chl-*a* m$^{-3}$) as it is the easiest measure of phytoplankton biomass to obtain. Irradiance has been converted to μmol photons m$^{-2}$ s$^{-1}$ from the energy units (W m$^{-2}$) used in most of the original reports. For the typical incubation light source used in these experiments, energy units can be converted to photosynthetically active photons (400–700 nm) using the approximation 1 W ≈ 4.6 μmol s$^{-1}$. Incubation light sources changed several times[17] but we did not correct the energy-quanta conversion or the photosynthetic efficiency[23] for changes in the spectrum of the light source.

### Statistical analysis

We classified each PI curve according to whether the data exhibit just the linear response, a saturating response, or a saturating response extending into photoinhibition (Supplemental methods). We used only PI curves classified as exhibiting photoinhibition for subsequent analyses since a comparison of photoinhibition models on data that do not exhibit photoinhibition would not be informative. Maximum likelihood estimation using nonlinear optimization was used to estimate the parameters for each model in Table 1 as documented in our R package piCurve[24].

We computed four statistics for each PI curve and each model: the root mean squared error (RMSE), the Akaike Information Criterion modified for small sample sizes (AICc), the Bayesian Information Criterion (BIC) which has a larger penalty for the number of observations and parameters compared to the AICc, and the coefficient of determination adjusted for the number of parameters ($R^2_{adj}$) (Supplementary Methods). We scored each model using the median RMSE over all PI curves. The RMSE was computed in two ways: on all PI data and only on the PI data with saturating or photoinhibiting irradiance, to emphasize performance of the photoinhibition models. A model with one more parameter than another model is generally expected to have a lower RMSE. We placed models into groups according to their number of parameters (3, 4, or 5) and identified the model in each group with the smallest median RMSE. Out of this set of three models, we identified the best model as the one with the smallest BIC on the most PI curves. The AICc was used as a check on the assessment using BIC. We winnowed the full set of models to the model with lowest RMSE for each number of parameters before ranking based on BIC scores because ranking results vary according to the set of models being compared. The difference in RMSE for each model was compared to the best model with the same number of parameters using a *t*-test. The adjusted coefficient of determination, $R^2_{adj}$, was recorded for each model, but it was not used to evaluate models.

We used a non-dimensionalized version of Eq. (3) to assist in the interpretation of $I_\alpha = P_{max}/\alpha$ and $I_\beta = P_{max}/\beta$:

$$\hat{P} = \tanh(\hat{I}) \tanh\left(\left(\frac{I_\beta}{I_\alpha \hat{I}}\right)^\gamma\right), \quad (4)$$

where $\hat{P} = \frac{P}{P_{max}}$, $\hat{I} = \frac{I}{I_\alpha}$ and $\gamma = \cosh^2(1)$. With this scaling, the irradiance parameters, $I_\alpha$ and $I_\beta$, can be interpreted by observing that $\hat{P}(I = I_\alpha) = \hat{P}(I = I_\beta) \approx \tanh(1) \approx 0.76$. Thus, the irradiance parameters $I_\alpha$ and $I_\beta$ identify the irradiance at which photosynthetic rate is about 76% of its maximum. This is analogous to the half-saturation constant of Michaelis–Menten models but with a different ratio; for example, in Eq. (1c), $I_\alpha$ is the irradiance at which P is half of $P_{max}$. Other thresholds can be found to more narrowly approximate the interval of the plateau; for example, $P(1.5\ I_\alpha) = P(I_\beta^\gamma/1.5) \approx 0.9\ P_{max}$.

### Additional analyses described in supplement

Parameters in all models have similar interpretations (initial slope of the PI curve, α; maximum photosynthetic rate $P_{max}$; saturating irradiance, $I_\alpha$; photoinhibition parameter β and irradiance $I_\beta$) but since the algebraic form of each equation differs, numerical values obtained from the same data will differ. We used PI curves simulated from the distribution of observed parameters to compare parameter values estimated from different models. Algebraic equivalences between similarly named parameters in different models are summarized in Table S2. Each PI curve with photoinhibition was classified according to whether it exhibited a plateau if the model with the best fit had a plateau. Two of our models (Amirian, exp-tanh) use a fixed value for the shape parameter ($\gamma = \cosh^2(1)$) which is motivated algebraically (Eq. S1) and justified statistically. We performed a statistical analysis for light-saturating models (Table S1) on PI curves that did not exhibit photoinhibition.

## Results

PI curves in our database were obtained from samples spanning the four seasons and more than 60° of latitude and thus are representative of samples from many oceanic conditions (Fig. S1). About half (1808 of 3641) of the PI curves in our database exhibited photoinhibition. The models we introduced in this paper can capture a plateau in the PI curve, while most previously introduced models have an absolute maximum photosynthetic rate at a single irradiance when photoinhibition is present. The vast majority (87% or 1574 scored by root mean squared error (RMSE)) of PI curves with photoinhibition were better represented by a model with a plateau compared to any of the models without a plateau (Table S3). A complete set of PI parameters with error estimates for each model and each PI curve is provided in supplemental data.

The Amirian model (Eq. 3 with $\gamma = \cosh^2(1)$) had the smallest mean RMSE indicating it had the best fit to the data on average among all models with four parameters (Table 2, Fig. S3, *t*-tests, $p < 0.001$) across our database of 1808 PI curves with photoinhibition.

**Table 2 | Statistical summary of 1808 photoinhibition model fits, ordered by increasing mean root mean squared error (RMSE)**

| Model | *p* | All irradiance | | | Excluding $I < I_\alpha$ | | |
|---|---|---|---|---|---|---|---|
| | | Median $R^2_{adj}$ (%) | Mean RMSE | Mean rRMSE (%) | Median $R^2_{adj}$ (%) | Mean RMSE | Mean rRMSE (%) |
| **Tanh-tanh-γ** | **5** | **96.5** | **0.144** | **−5.860** | **64.6** | **0.161** | **−8.080** |
| Exp-tanh-γ | 5 | 96.4 | 0.146 | −4.620 | 64.2 | 0.163 | −7.360 |
| Fasham & Platt[21] | 5 | 96.3 | 0.148 | −3.340 | 57.7 | 0.172 | −1.590 |
| **Amirian** | **4** | **96.1** | **0.151** | **0.000** | **59.0** | **0.175** | **0.000** |
| Exp−tanh | 4 | 95.9 | 0.156 | 3.320 | 58.5 | 0.180 | 2.920 |
| Smith[32a] | 4 | 95.7 | 0.161 | 7.360 | 52.7 | 0.194 | 14.72 |
| Bannister[26a] | 5 | 95.4 | 0.161 | 8.780 | 50.8 | 0.193 | 15.98 |
| Prioul & Chartier[25a] | 5 | 95.4 | 0.166 | 9.250 | 47.4 | 0.198 | 16.06 |
| Exp-exp | 4 | 95.6 | 0.162 | 9.170 | 56.4 | 0.193 | 14.67 |
| Neale & Richerson[20] | 4 | 95.5 | 0.164 | 10.62 | 48.7 | 0.198 | 18.52 |
| Platt et al.[16] | 4 | 95.4 | 0.169 | 13.29 | 47.5 | 0.204 | 21.28 |
| Tanh−exp | 4 | 95.4 | 0.164 | 10.69 | 52.1 | 0.197 | 17.92 |
| Blackman[33a] | 4 | 94.6 | 0.180 | 27.96 | 40.1 | 0.212 | 33.46 |
| Baly[9a] | 4 | 94.3 | 0.186 | 25.80 | 44.1 | 0.228 | 36.50 |
| Peeters & Eilers[28] | 4 | 93.5 | 0.195 | 34.64 | 37.8 | 0.243 | 49.45 |
| **Steele[15]** | **3** | **89.3** | **0.248** | **74.03** | **22.4** | **0.279** | **70.25** |

The number of parameters is p. Statistics are median adjusted $R^2$ (%), mean RMSE, mean relative RMSE (rRMSE) computed over all irradiances and, in the last three columns, excluding data points in the light limited region ($I < I_\alpha$). The best model for each number of parameters is highlighted in bold. Positive rRMSE indicates a higher RMSE than the Amirian model, while negative values show a reduction—typically seen in models with additional parameters.

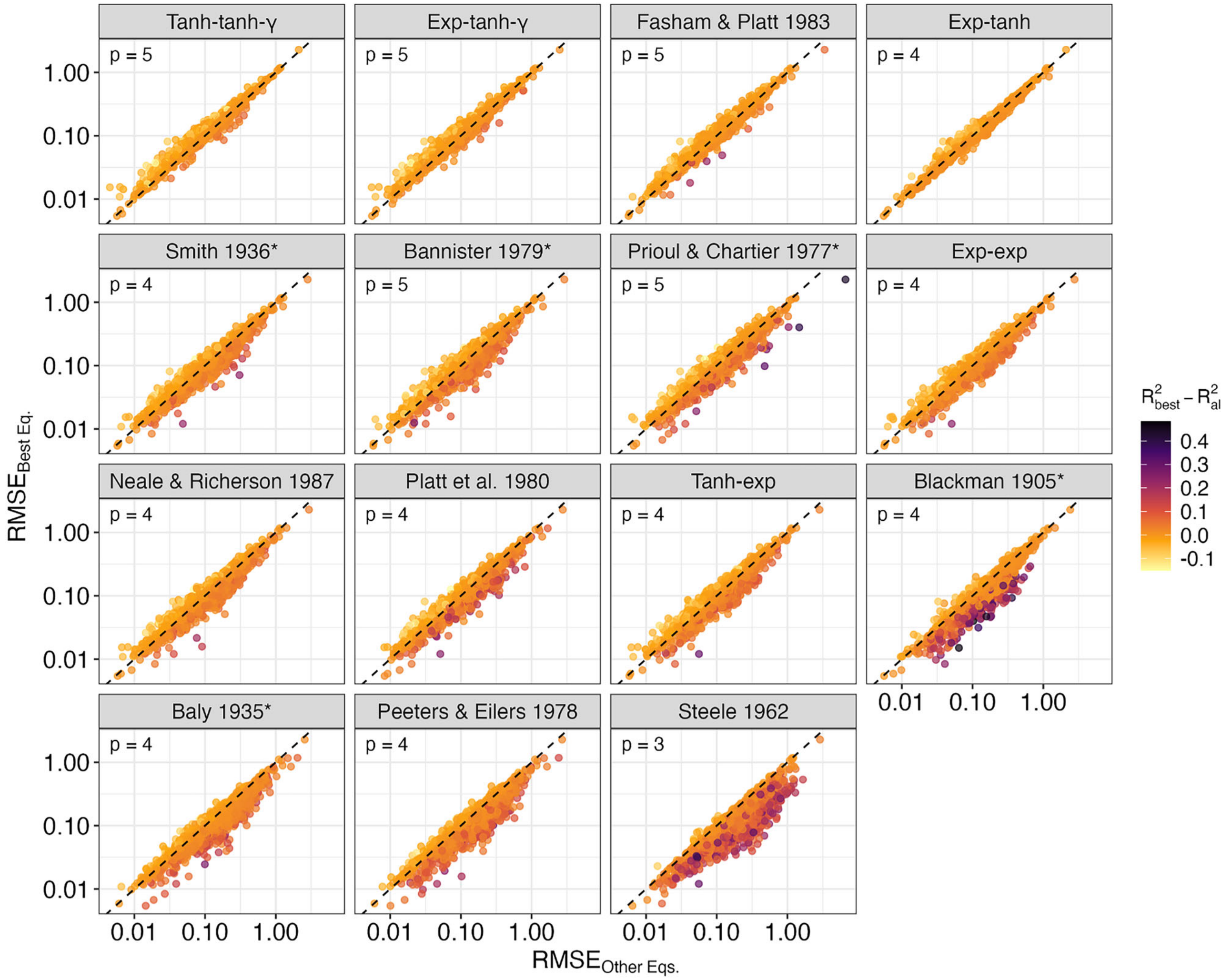

**Fig. 2 | Model performance (RMSE, log scale) of the best four-parameter model (Amirian) compared to all other models (indicated in the header for each panel).** Each point corresponds to a separate PI curve. Point color corresponds to the difference in $R^2_{adj}$ between Amirian model and the comparison model. The 1:1 line is shown as a black dotted line. Models are defined in Table 1, and $p$ in each panel indicates the number of parameters in the corresponding model.

A comparison of RMSE and $R^2_{adj}$ for the Amirian model vs. other models illustrates the variation in gains in model performance across many PI curves (Figs. 2, 3 and S9). Our model achieved up to a 45% improvement in $R^2_{adj}$ relative to existing models. Models with an additional shape parameter showed up to a 10% improvement in $R^2_{adj}$ compared to Amirian model, but these gains were statistically insignificant ($t$-test, $p > 0.05$) and therefore those models were not judged as superior. Although the median $R^2_{adj}$ for most models was high (>0.9), this metric is often dominated by the light-limited region of the PI curve, which contains most of the observations. As such, $R^2_{adj}$ alone does not reliably reflect differences in model performance within the photoinhibition-dominated region. When adjusted $R^2$ and RMSE were computed excluding the light limited region of each PI curve, the Amirian model improved the median adjusted $R^2$ by 6.3% to 36.6% and reduced RMSE by 14.7% to 70.2% compared to existing models. Notably, it achieved a 21.3% improvement in RMSE over the most commonly used model (Eq. 2), and outperformed models with greater complexity, reducing RMSE by 16% relative to the Prioul & Chartier[25] and Bannister[26] models, which both have one additional parameter (Table 2).

Among the models with five parameters, the Amirian model with an extra shape parameter (tanh-tanh-γ) had the smallest root mean squared error (RMSE, Table 2). The estimated mean value for the shape parameter in this model was $\gamma = 2.30 \pm 0.09$ (95% CI) which is statistically indistinguishable from the shape parameter used ($\gamma = \cosh^2(1) \approx 2.38$) in the four-parameter Amirian model. Serendipitously, this value for γ yields an especially simple interpretation for β as the magnitude of the slope of photosynthetic rate at $I = I_\beta$ (see Supplement, Fig. S2). The four parameter Amirian model (with fixed shape parameter) had the smallest BIC and AICc for the most PI curves compared to the 3 and 5 parameter models with the smallest RMSE (Table 3). An additional shape parameter did not significantly enhance any model's ability to capture the photoinhibition part of the data, relative to the Amirian model.

PI curve parameters are distributed approximately log-normally and positively correlated with a left skew in $I_\alpha$ and $I_\beta$ (Figs. S4 and S5). The distribution of photoinhibition parameters (β, $I_\beta$) shows the range of photoinhibition rates and irradiances at which photoinhibition becomes quantitatively important across the PI curves in our database (Fig. S4, Table S4). The photoinhibition rate β is frequently less than 10% of the photosynthetic efficiency α indicating that the decline in photosynthetic rate at high irradiance is more gradual than the increase at low irradiance. Five PI curves drawn using the Amirian model illustrate the range of shapes typically observed in the data (Fig. 4). A plateau is noticeable when the ratio $I_\beta/I_\alpha$ is larger than 8, which occurred in 82% of PI curves with photoinhibition. This ratio is symmetrically distributed around 12, indicating that a substantial plateau is typically present in the data (Fig. S4).

The dark respiration parameter, R, was not statistically different from 0 for most models most of the time (50-70% of PI curves, except Steele 1962, exp-tanh, and exp-tanh-γ, not shown). Following past practice[16], we do not interpret patterns in this parameter but do report it in our results.

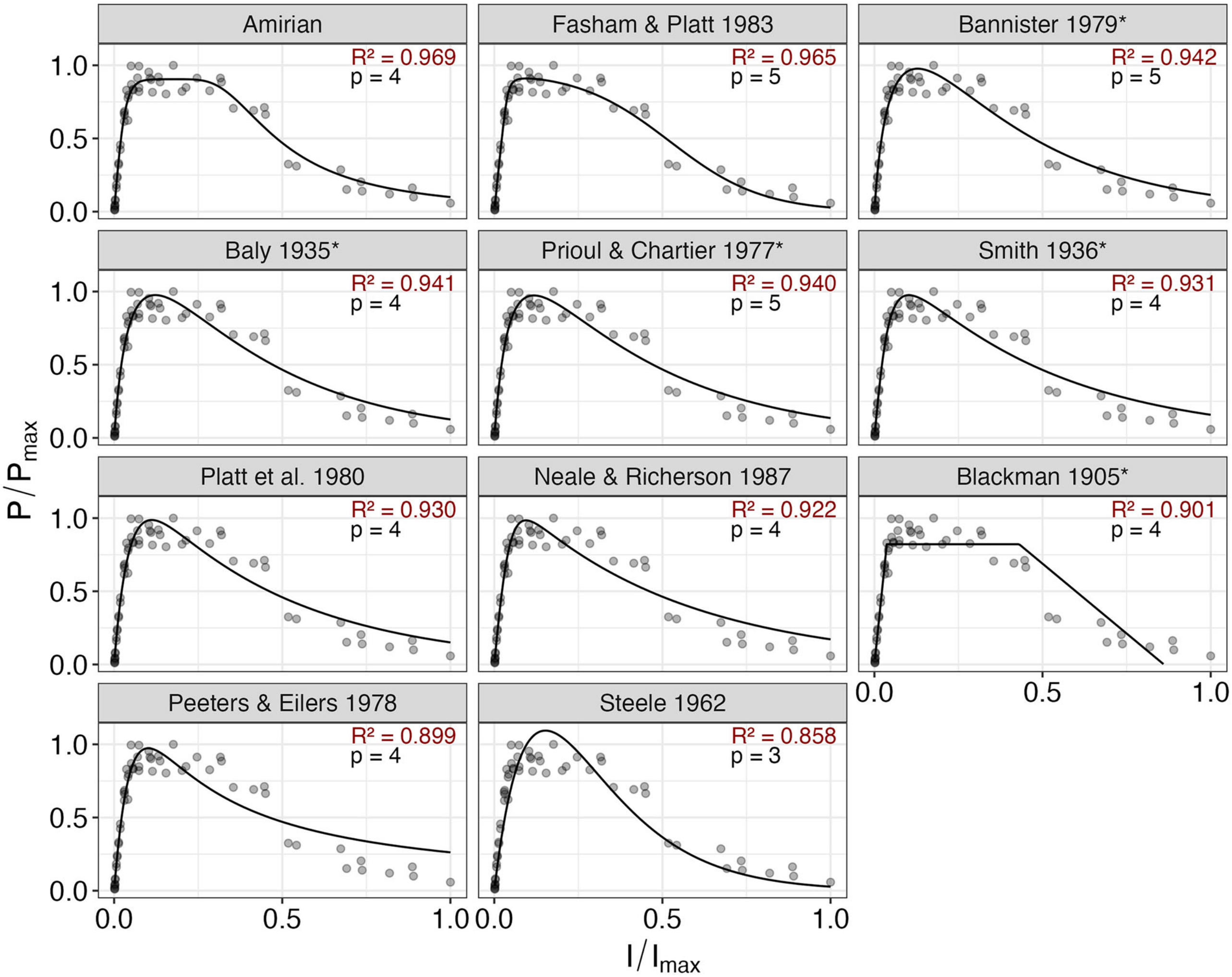

**Fig. 3 | Comparison of the Amirian model (Eq. 3, with $\gamma = \cosh^2(1)$) with all previously published models (Table 1), applied to a photosynthesis–irradiance data sample[12].** Models (black lines) are sorted from best to worst based on $R^2$ performance (shown in red), and the value of $p$ indicates the number of parameters in each model.

**Table 3 | Performance of the three models with lowest median RMSE for each number of parameters, p, assessed as the number of times (and percentage) each model had the lowest BIC and AICc over all PI curves with photoinhibition**

| Model | *p* | Number of times (and percent) each model was favored by | |
|---|---|---|---|
| | | BIC (%) | AICc (%) |
| Amirian | 4 | 1009 (56) | 983 (54) |
| tanh-tanh-γ | 5 | 689 (38) | 701 (39) |
| Steele[15] | 3 | 110 (6) | 124 (7) |

Our analysis of PI curves not exhibiting photoinhibition echoed the results of a previous analysis[11] (Supplemental results, Table S5) and enabled us to compare photosynthetic efficiency and capacity across PI curves with and without photoinhibition. We observed a statistically significant reduction in median $P_{max}$ (44%) and a smaller reduction in median α (9%) in PI curves when photoinhibition was present compared to PI curves that did not exhibit photoinhibition (Fig. S6).

## Discussion

Photosynthesis–irradiance (PI) curves are valuable tools for summarizing phytoplankton photosynthetic performance, computing water-column integrated primary production, and parameterizing phytoplankton growth in large-scale ecosystem models. Jassby and Platt[11] analyzed many models over sub-saturating and saturating irradiances and recommended the now widely used hyperbolic tangent functional form (Eq. 1a). No similar comparative analysis for photoinhibition models has been presented in the literature, although many models have been introduced. In our database, about half of all PI curves exhibited photoinhibition and of these more than three-quarters included a plateau where photosynthetic rate was approximately constant over a range of irradiances (Table S3). Most models of photoinhibition are qualitatively inadequate as they do not include this plateau. Photoinhibition models that do not capture a plateau and PI models that ignore photoinhibition will typically lead to biased estimates of parameter values as the shape of those functions do not match the shape of the data (Fig. 3). Here we introduced a new model with a plateau, evaluated it against commonly used models, and showed that our new model is quantitatively and qualitatively superior to existing models (Figs. 2, 3, S7, and S9), making it a reliable model used in estimating primary production in the water column[27].

Our analysis supports the use of a new parsimonious photoinhibition model (Amirian model, Table 1) that only adds one parameter to the commonly used photosynthesis–irradiance model (Eq. 1a), for a total of four parameters (α, $P_{max}$, β, and R). In the absence of photoinhibition ($\beta = 0$), the Amirian model simplifies to the most widely used model without photoinhibition (Eq. 1a, Fig. S2). The plateau in our model separates the light-limited and light-saturated regions of the curve from the photoinhibition regions, so that the interpretation of parameters $P_{max}$ and α are not affected by the presence of photoinhibition. Adopting our model will allow researchers to avoid changes in parameter interpretation between commonly used models with (Eq. 2) and without (Eq. 1a) photoinhibition. This mathematical consistency minimizes under- or over-estimation of PI parameters caused by switching between models with and without photoinhibition, reducing uncertainty in parameter estimates, which is crucial for large-scale primary production calculations.

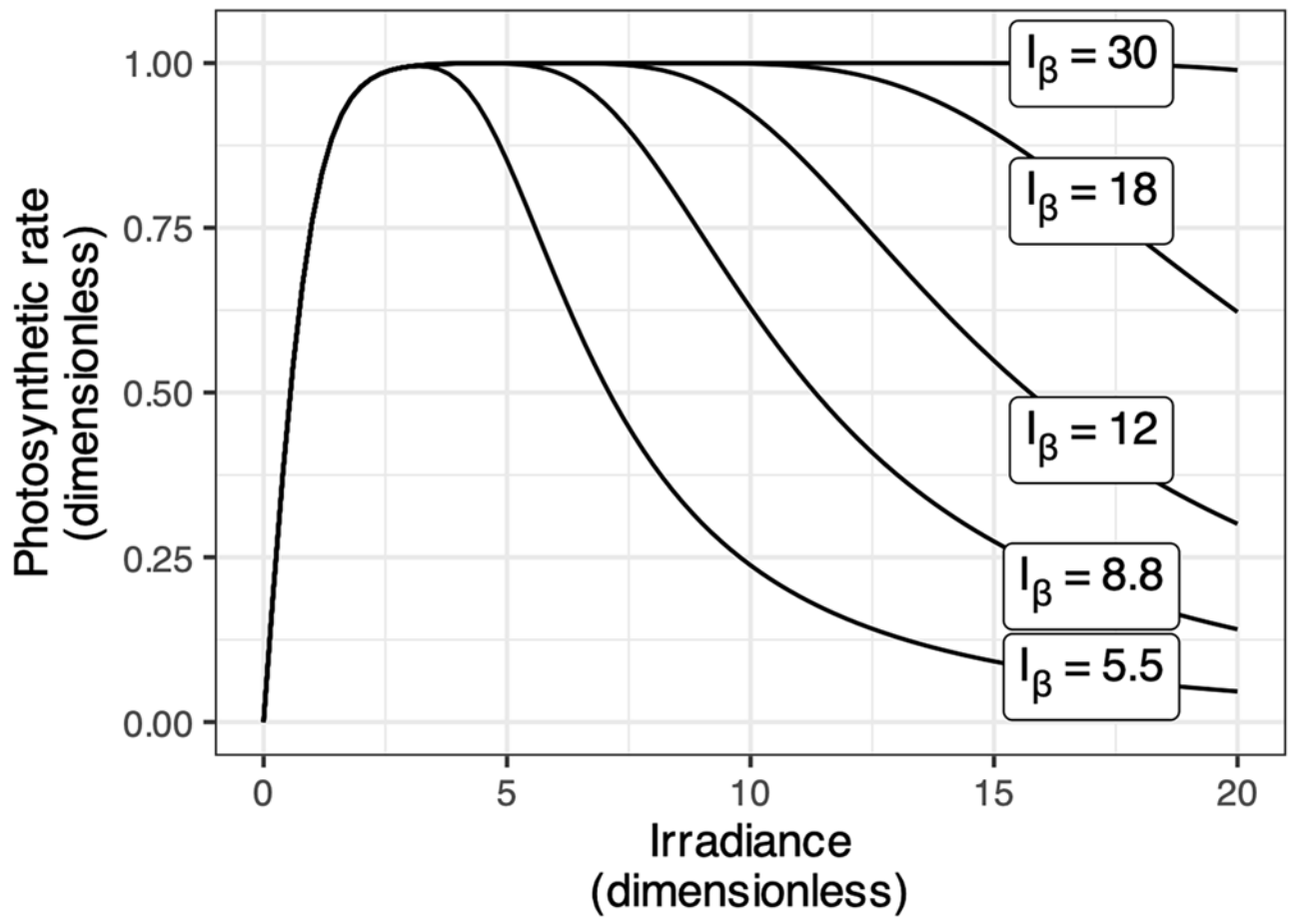

**Fig. 4 | Dimensionless Amirian model (Eq. 3, $P_{max} = 1$, $\alpha = 1$, $I_\alpha = 1$) representing the range of observed PI curves with photoinhibition, illustrated using 5 quantiles (5%, 25%, 50%, 75%, 95%) of non-dimensionalized $I_\beta$ and $\beta = 1/I_\beta$ values.** The dimensionless value of $P(I = I_\beta)$ is $\tanh(1) \approx 0.76$.

Analysis of empirical and simulated datasets reveals up to 40% differences in the α parameter (the slope of the PI curve at low irradiances) between the Jassby & Platt[11] and Platt et al.[16] models (Table S6).

Our new photoinhibition model significantly improves the characterization of photosynthesis–irradiance (PI) curves compared to existing formulations. Across 1808 PI curves, it consistently outperformed all other four-parameter models, achieving the lowest mean and median RMSE (Table 2), with statistically significant improvements confirmed via pairwise *t*-tests (Fig. S3, $p < 0.001$). The Amirian model showed up to a 45% improvement in $R^2_{adj}$ relative to other models (Fig. 2). Crucially, when the analysis focused on the light-saturated and photoinhibited portion of the curve—where our model is intended to bring the most value—it reduced RMSE by 21% compared to the widely used Platt et al.[16] model, and by as much as 49% compared to the Peeters & Eilers[28] formulation. Even against more complex models with additional parameters, such as Prioul & Chartier[25] and Bannister[26], our model reduced RMSE by 16%. These substantial improvements demonstrate that the Amirian model not only maintains parsimony, but also provides a more accurate and robust framework for modeling photoinhibition, especially in cases where a plateau in photosynthetic rates is observed (Fig. 3, S7, and S9).

The photoinhibition parameter, β, in the Amirian model is the magnitude of the slope of the PI curve at the photoinhibition irradiance, $I_\beta$ (Fig. S2), echoing the photosynthetic efficiency, α, which is the slope at zero irradiance. The photoinhibition irradiance, $I_\beta = P_{max}/\beta$, describes the irradiance at which photoinhibition becomes important, echoing the light-saturation irradiance. In the widely used photoinhibition model (Eq. 2), β describes a photodamage process starting at zero irradiance with no easy interpretation for $I_\beta$.

Previous work has suggested that describing photoinhibition requires two additional parameters: one for the irradiance where photoinhibition begins and another for the rate of photoinhibition[16,20]. It is difficult to estimate the irradiance at the start of photoinhibition in models defined with sharp thresholds because experiments usually have large gaps between irradiance treatments in this part of the curve and small steps in irradiance are needed to identify the threshold irradiance. As a result, estimates of the irradiance where photoinhibition begins are generally statistically underpowered. In our evaluation, we found that some models with a fifth parameter can capture more of the variability in observations (smaller RMSE, Table 2). We rejected these five parameter models on statistical grounds that penalize models with more parameters (Table 3).

Our new model captures qualitative features of PI curves, notably the plateau in photosynthetic rate when photoinhibition observed in the experiment and a symmetry between the increase and decrease in photosynthetic rate with increasing irradiance. One existing photoinhibition model captures the plateau in photosynthetic rate without a difficult to estimate transition irradiance[21]. This model has an extra shape parameter compared to the Amirian model and does not simplify to a widely used light-saturating model in the absence of photoinhibition. We rejected this more complex model as not parsimonious. The root mean squared errors for this model are similar to the results for our new model and larger than errors for other models with five parameters (Table 2). The development of the Fasham & Platt[21] model is grounded in a mechanistic description of photodamage and is linked to interpretable rate parameters, but despite these features it has not been widely adopted in the literature. The Eilers & Peeters[29] model incorporates symmetric increases and decreases in photosynthetic rate on a log irradiance scale (their Fig. 3[29]). Our reciprocal irradiance formulation allows our model to capture this symmetry. All six of our new models (Table 1) have this property, but the hyperbolic tangent function is the best model for photoinhibition, mirroring the shape found for light saturation[11] in the decrease in photosynthetic rate with photoinhibition.

The correlation between α, $P_{max}$, and β is dominated by a positive, log-linear relationship (Fig. S5). Multiple processes appear to be responsible for the correlation and residual variation. One hypothesis is that cells acclimated to low light have high photosynthetic efficiency, relatively low photosynthetic capacity, and show high levels of photoinhibition because of susceptibility to photodamage, but this is not the dominant pattern in the data. Photosynthetic capacity and efficiency are sometimes uncorrelated as a result of photoacclimation and sometimes positively correlated due to changes in the metabolic processing of photosynthetic reductant[30]. More work is needed to explore the causes of variation in correlations between β and the other photosynthetic parameters.

Photoinhibition is frequently observed in photosynthesis–irradiance curves and models that represent the phenomenon poorly will introduce bias into PI parameters and predictions of photosynthetic rate. Many processes contribute to the plateau and decrease in photosynthetic rate with increasing irradiance, including changes to the antenna and photoprotective pigment cellular content, non-photochemical quenching, acclimation to changing growth conditions and resources, and photodamage and repair of photosystem II. Several models have been developed to account for the some of these processes[12,14,21,29,31]. These models are valuable contributions to the study of these mechanisms, but the complexity of photoinhibition means that they are generally incomplete descriptions of the observed PI data. In our view the difficulty in accurately describing the numerous complex mechanisms means there is considerable value in the simple phenomenological parameterization developed here.

## Data availability

Photosynthesis–irradiance data, fitted model parameters and other statistics are deposited at zenodo (https://doi.org/10.5281/zenodo.16748102). Data in Fig. 1 are identified as PI curves PI002600 and PI000375 in this database, and PI002788 for Figs. 3 and S9.

## Code availability

Code for the R analysis system is available on GitHub in the package Mohammad-Amirian/piCurve[24].

## Acknowledgements

We are indebted to many scientists at the Bedford Institute of Oceanography, Fisheries and Oceans Canada in Dartmouth, Nova Scotia, Canada, who over five decades have collected the data we analyzed. Financial support for this work was provided by the Simons Foundation CBIOMES (Computational Biogeochemical Modeling of Marine Ecosystems) collaboration (Award 549935 to A.J.I. and 986772 to Z.V.F.).

## Author contributions

Mohammad Amirian developed the model and performed the statistical analysis, assembled and quality controlled the data, wrote and edited the paper. Emmanuel Devred contributed data and edited the paper. Zoe Finkel contributed to the analysis, wrote and edited the paper. Andrew Irwin contributed to model design and statistical analysis, wrote and edited the paper.

## Competing interests

The authors declare no competing interests.

## Additional information

**Supplementary information** The online version contains supplementary material available at https://doi.org/10.1038/s43247-025-02686-3.

**Correspondence** and requests for materials should be addressed to Mohammad M. Amirian.

**Peer review information** *Communications Earth & Environment* thanks the anonymous reviewer(s) for their contribution to the peer review of this work. Primary Handling Editors: Dania Albini and Alice Drinkwater. [A peer review file is available].